\documentclass[a4paper,twocolumn,%tightenlines,
english,aps,pre,floatfix,superscriptaddress,showpacs]{revtex4-1}
\usepackage[T1]{fontenc}
\usepackage[latin1]{inputenc}
\usepackage{amsmath}
\usepackage{babel}
\usepackage{graphics}
\usepackage{amssymb}


\begin{document}
\title
{Renormalization group treatment of rigidity percolation}

\author {R. B. \surname{Stinchcombe}}
\email{r.stinchcombe1@physics.ox.ac.uk}
\affiliation{Rudolf Peierls Centre for Theoretical Physics, University of
Oxford, 1 Keble Road, Oxford OX1 3NP, United Kingdom}
\author {M. F.\surname{Thorpe}}
\email{mft@asu.edu}
\affiliation{Rudolf Peierls Centre for Theoretical Physics, University of
Oxford, 1 Keble Road, Oxford OX1 3NP, United Kingdom}
\affiliation{Physics Department, Arizona State University, Tempe, AZ 85287}

\date{\today}

\begin{abstract}
Renormalization group calculations 
are used to give exact 
solutions for rigidity 
percolation on hierarchical lattices. 
Algebraic scaling transformations 
for a simple example in two dimensions 
produce a transition of
second order, with an unstable critical point
and associated scaling laws. 
Values are provided 
for the order parameter 
exponent 
$\beta = 0.0775$ associated with
the spanning rigid cluster and also 
for 
$d \nu = 3.533$ which is associated with an anomalous lattice dimension $d$ and the divergence in the correlation length
near the transition.  In addition we argue that the number of floppy modes $F$ plays the role of a free energy and hence find the exponent $\alpha$ 
and establish hyperscaling. 
The exact analytical procedures demonstrated on the chosen example readily generalize to wider classes of hierarchical lattice.
\end{abstract}
\pacs{05.70.Jk, 05.70.Fh, 62.20.-x }
%05.70.Jk	Critical point phenomena
%05.70.Fh	Phase transitions: general studies
%62.20.-x	Mechanical properties of solids
\maketitle
%\tightenlines

In this letter we re-visit rigidity percolation on a lattice and show for the first time how renormalization group 
calculations can be 
exactly performed on 
particular bond-diluted hierarchical lattices 
in two dimensions 
and  show that the transition is second order. 
This is in contrast to the only other exact solution known for the rigidity transition, on Cayley tree networks which is first order~\cite{Leath, Duxbury}. 

Phase transitions associated with rigidity have experimental importance in the elastic behavior in chalcogenide glasses~\cite{Thorpeg}, in protein unfolding~\cite{Thorpep} and in jamming in granular materials~\cite{Nagel}. Rigidity 
percolation  is similar conceptually to the more familiar connectivity percolation~\cite{Stauffer, RBS}, except that instead of demanding a connected pathway
across the sample, the more stringent condition  that the connected  pathway is also{\it{ rigid}} is required.

Rigidity percolation  on networks has been studied since 1984  when the concept was first introduced and a mean field description proved remarkably accurate~\cite{Feng1, Feng2}, 
except very close to 
the phase transition. 
Subsequent work  has been mainly numerical~\cite{Jacobs1,Jacobs2}. The 
associated  
rigidity  phase transition has been most extensively 
investigated 
on the triangular network in two dimensions where numerical studies 
(using the pebble game algorithm 
outlined below) 
show that the transition is second order 
and 
described by critical exponents $\beta = 0.18 \pm 0.02 $  and  $d\nu = 2.42 \pm 0.12$ that are distinct from those of connectivity percolation ($\beta = {5/ 36} = 0.139$  and  $d\nu = {8 / 3} = 2.667$).

Results in three dimensions using the pebble game algorithm ~\cite{Chubynsky} strongly suggest that the rigidity transition is first order on
a bond diluted face centered cubic lattice, whereas if angular forces are included whenever two adjacent edges are present, the transition is second order. This is quite different from connectivity percolation where the transition is always second order in three dimension~\cite{Stauffer}. Further information comes from Cayley tree networks where connectivity percolation is second order, whereas rigidity percolation (from a rigid busbar) shows a strongly first order transition~\cite{Duxbury}.

\begin{figure}
{\centering \resizebox*{2.3in}{!}{\includegraphics*{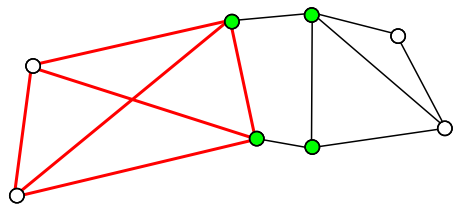}}}
\caption{(Color online) An isolated piece of network is used shows an isostatic (unstressed) rigid region with edges shown by thin lines 
and an overconstrained (stressed) rigid region containing  edges shown as thick lines. At the flexible hinges, shown as solid  circles, angular motion is possible.} 
\label{fig:a}
\end{figure}

Some characteristics of networks related to rigidity are illustrated in Fig.~\ref{fig:a}. Such particular network realizations are elucidated by exact counting procedures~\cite{Maxwell, Laman} such as the Maxwell count (used below) and in the pebble game algorithm. Both 
balance constraints against degrees of freedom.
The latter finds the 
rigid clusters and the flexible joints between them
and 
also determines redundant bonds in overconstrained regions, as
illustrated 
in  Fig.~\ref{fig:a}.

\begin{figure}
{\centering \resizebox*{2.5in}{!}{\includegraphics*{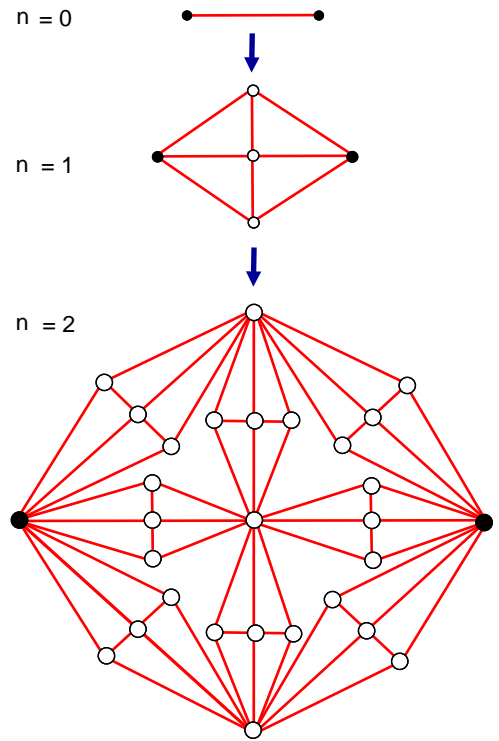}}}
\caption{(Color online) Each generation of the undiluted hierarchical lattice is labeled by an index $n$ and $n=0 ,1, 2 $ shown here starting at the top.}
\label{fig:b}
\end{figure}

In this letter 
we show how exact calculations can be 
performed on 
hierarchical networks in two dimensions, taking for detailed discussion the 
Berker lattice~\cite{Berker, RBS2}
shown in Fig.~\ref{fig:b}. 
When diluted, this network is one of the simplest which captures fundamental generic features for rigidity percolation. 
Each generation is obtained by decoration of the previous generation, creating an infinite sequence that can lead to singularities and a phase transition. An exact set of equations 
can 
be written down, relating quantities associated with generations $n+1$ and $n$,  which can be solved at all bond concentrations $p$ by iteration. Most importantly the stable and unstable fixed points can be found and the structure of the rigidity phase transition can be described  by 
the scaling behavior obtained by 
expanding about the unstable fixed point.

It is 
instructive to do a Maxwell count~\cite{Thorpe} on the first three generations of the Berker lattice shown in Fig.~\ref{fig:b}. The number of floppy modes $F$ is given by the difference in the number of degrees of freedom $2V$,  associated with the number of vertices $V$, 
and 
the number of constraints which are associated with the 
number $E$ of edges. 
However, in general not all the edges are independent constraints and so $E$ must be corrected by the number $R$ of redundant edges so that

\begin{equation}
F=2V-E +R .
\label{eq:aa}
\end{equation}
The number of floppy modes  $F$ in Eq.~(\ref{eq:aa}) contains the  3 
rigid body motions 
in two dimensions 
(two translations and one rotation) that become insignificant in the limit of a very large number of edges.  For the top panel in Fig.~\ref{fig:b}, and  ignoring $R$ for the moment, $F-3 =2\cdot2 -1-3 = 0$ while for the second and third panels $F-3 =2\cdot5-8-3=-1$ and $F-3 =2\cdot29 - 64 -3=-9$ respectively, where the negative numbers signify that not all the bonds are independent in these rigid diagrams, and we have removed the 3 macroscopic floppy modes on the left hand side of the count. Therefore there is a single redundant edge  in the second panel and 9 redundant edges in the third panel of Fig.~\ref{fig:b} (one for each of the eight replications of the second panel, plus one new one). By removing edges randomly from the third panel
(i.e. bond diluting), 
first the redundancy is reduced and eventually there is no rigid path between the two solid vertices, and rigidity is lost. Note that the Maxwell count for Fig.~\ref{fig:a} gives $F-3= 2\cdot8-13-3= 0$ (again removing the three floppy modes), but as there is one floppy mode associated with the solid vertices, there is also one redundant edge associated with the heavier solid edges in the left side of the diagram.

The number of edges $E$ or more precisely  bonds $N_n^{b}(p)$, and  the number of vertices $V$  or sites  $N_n^{s} (p)$
 in the $n^{th}$ generation becomes, for the undiluted case ($p=1$)  shown in  Fig.~\ref{fig:b} is

\begin{eqnarray}
E &=& N_n^{b} (1) = 8^n,   
\nonumber \\
V &=& N_n^{s} (1) = (3.8^n+11)/7 .
\label{eq:aaa}
\end{eqnarray}
An important quantity is the mean coordination defined by $\langle{r} \rangle = 2  E/V $ which tends to an asymptotic value $\langle{r} \rangle = 14/3 = 4.667$ for the undiluted lattice. It is important that this quantity be above 4, which is the mean field value of the mean coordination needed for rigidity in two dimensions~\cite{Feng2, Jacobs2}. The number of redundant edges is  $(8^n - 1)/7$ so that the fraction of redundant edges for large $n$ approaches $1/7 = 14.3\%$ in the undiluted lattice. For the triangular lattice, this fraction is even higher at $1/3 = 33.3\%$.

\begin{figure}
{\centering \resizebox*{3in}{!}{\includegraphics*{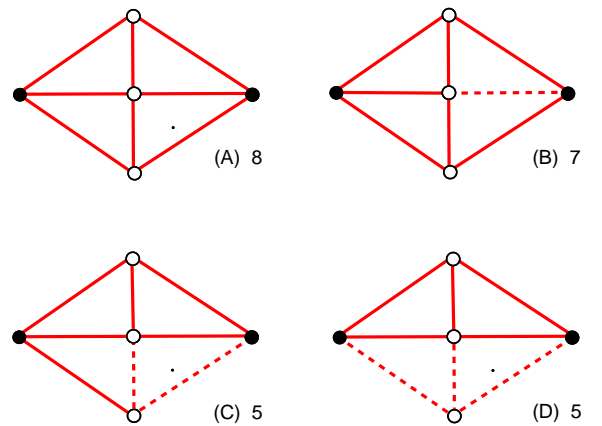}}}
\caption{(Color online) Showing the four distinct types of graphs that lead to a rigid connection between the two solid 
circles. Edges present are shown as solid lines and missing edges as dashed lines. 
Here (A) has all eight bonds present and is rigid with one redundant edge and has probability $p^8$, (B) has any single edge missing and has probability $8p^7(1-p)$, (C) has any pair of edges missing from the three lower (shown) or 
the three upper ones and has probability $6p^6(1-p)^2$ and (D) has a triple of edges missing either from the lower or upper part of the graph and  has probability $2p^5(1-p)^3$. The number of edges in the rigid cluster is indicated by the number under each graph.
All other graphs (not shown)  do not rigidly connect the two solid circles.  } 
\label{fig:c}
\end{figure}

For the diluted case an 
bond is present with probability $p$ 
(concentration) 
and absent with probability $1-p$,  so the probability of the two solid dots being rigidly connected in the second panel of  Fig.~\ref{fig:b} using the weights from  Fig.~\ref{fig:c} is $p' = p^8 +8p^7(1-p) + 6p^6(1-p)^2 + 2p^5(1-p)^3 = 2p^5+2p^7-3p^8$. 
This leads to the relationship between the probabilities $p_{n+1}, p_{n}$ of rigidity percolating in successive  generations $n+1, n$:

\begin{equation}
p_{n+1} = 2p_n^5+2p_n^7-3p_n^8
\label{eq:a}
\end{equation}
(with $p_0=p$). 
The 
fixed points $p^*$ 
satisfying 
$p_{n+1}=p_n=p^*$  
are the
trivial stable fixed points at $p^*=0$ and $p^*=1$ and 
the
non-trivial unstable fixed point 
$p^*=0.9446
=p_c$. Close to this 
latter
fixed point,  Eq.~(\ref{eq:a}) can be linearized by differentiating to give 
$(p_{n+1} -p_c) = \lambda_1 (p_n - p_c)$  
where $\lambda_1 =  10p_c^{4} + 14p_c^{6}  -24p_c^{7} = 1.802$.

\begin{figure}
{\centering \resizebox*{3.3in}{!}{\includegraphics*{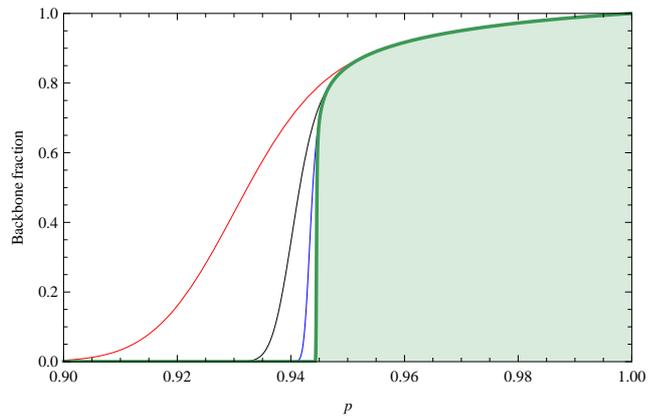}}}
\caption{(Color online) Showing the probability $P(p)$ that a bond that is present is also part of the rigid backbone as a function of 
the probability $p$ that an bond is present. The four curves shown are for the result of iterating Eq.~(\ref{eq:b}) out to $n =4,6,8,12 $ terms respectively as the curve steepens. The result for $n=12$ is  shown by the heavier line, and shows convergence on the scale of this plot to the singularity  at $p_c=0.9446$.  } 
\label{fig:d}
\end{figure}

Using the 
cluster 
probabilities and also the number of bonds in each rigid spanning cluster from  Fig.~\ref{fig:c}, 
we find from the mean number of bonds  that 
the probability 
$ P_{n+1}(p)$ of a bond belong to the percolating rigid cluster is given by the recurrence relation

\begin{equation}
P_{n+1}(p) = \frac{1}{4} [5p_{n}^4 + 13p_{n}^6  -14p_{n}^7] P_{n} (p)
\label{eq:b}
\end{equation}
(with $P_0(p)= p$). Near the unstable fixed point, $P_{n+1}(p) = \lambda_2 P_{n} (p)$ where $\lambda_2 =  
\frac{1}{4} [5p_c^{4} + 13p_c^{6}  -14p_c^{7}] = 0.9554$, 
showing that the probability of an bond being in the percolating cluster renormalizes  to zero at the phase transition as expected for a second order phase transition.  From  Fig.~\ref{fig:d} we can see how the singular behavior at the  phase transition develops as  $n$ 
increases: $n=12$
appears very close to 
giving
the 
full 
singularity. Near $p=1$, $P(p) = p[1-69{(1-p)^2}/4+...]$ where the first term is just the probability that an bond is present 
and the second term that at least 2 bonds must be removed to produce an bond that is present but not part of the rigid backbone, as indicated for example in panel (C) of Fig.~\ref{fig:c}.

Using the eigenvalues  $\lambda_1$, $\lambda_2$, and $\lambda_3 (=8)$ of the linearized scaling relationships for $p$, $P(p)$, and $N_n^{b}$ respectively we obtain exponents $\nu$, $\beta$, and {\it{fractal}} dimensionality $d$, 
from $\lambda_1 = b^{1/\nu}$,  $\lambda_2 = b^{-\beta/\nu}$, and $\lambda_3 = b^d$, where $b$ is the dilatation (length scaling) factor between successive generations of the hierarchical lattice. 
However, as is typical for such lattices, $b$ is ambiguous~\cite{RBS2}; so we quote only the values of exponents independent of $b$. These are $\beta = 0.0775$,  
which describes how the order parameter $P(p)$ goes to zero at the critical point, and the product $d\nu = 3.533$, which plays a role in hyperscaling which does apply here.

The question of hyperscaling involves
the critical exponent $\alpha$ that describes the fluctuations associated with the {\it{specific heat}} in the system near the phase transition. 
The exponent $\alpha$ 
is most easily calculated by differentiating the free energy 
twice 
with respect to 
the the bond concentration $p$, 
and hyperscaling also relates (when it applies) to the free energy.
But the question arises as to what is an appropriate free energy as
rigidity percolation 
is not a system described by a Hamiltonian. There is strong evidence, outlined below,  that the number of floppy modes given in Eq.~(\ref{eq:aa}) serves as the appropriate free energy 
for it.
It can be shown that the second derivative with respect to $p$ is positive definite. For connectivity percolation, the free energy can be found as the $s \rightarrow 1$ limit of the s-state Potts 
model~\cite{Fortuin} and in that case is equivalent to an appropriate version of 
Eq.~(\ref{eq:aa}) 
in which 
redundancy refers to loops or multiple pathways between two vertices
and the factor 2 is omitted. In this case a single floppy mode is associated with an isolated cluster, so the free energy is just the total number of isolated clusters and of course is an extensive quantity.
Finally for 
connectivity and rigidity, 
these forms of 
$F$ 
have 
been used as a free energy for percolation from a busbar onto a Cayley tree network~\cite{Duxbury}.  

\begin{figure}
{\centering \resizebox*{3.3in}{!}{\includegraphics*{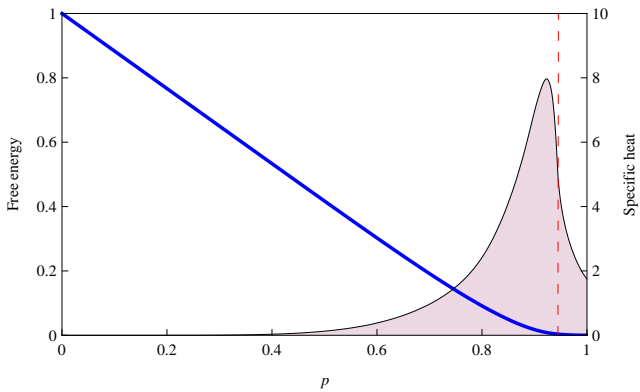}}}
\caption{(Color online) Showing the number of floppy modes $f(p)$ (heavy line and left scale )  and also its second derivative with respect to $p$ (thin line and right scale) which is the {\it{specific heat}}. The vertical dashed line marks the location of 
$p_c = 0.9446$ which is {\it{not}} at peak of the {\it{specific heat}}.} 
\label{fig:f}
\end{figure}

Rather than calculate the 
number $F$ of floppy modes 
directly, it is easier to calculate $R$ in Eq.~(\ref{eq:aa}) and hence determine $F$. If the number of redundant bonds at generation $n$ is $R_{n+1}(p)$, then 

\begin{equation}
R_{n+1}(p) = 8  R_{n} (p) + p_{n}^8 
\label{eq:d}
\end{equation}
(with $R_{0}(p) =0$). The factor 8 in  Eq.~(\ref{eq:d}) comes from the eight fold replication of any redundant bond from the previous generation (e.g. going from $n=1$ to $n=2$ in Fig.~\ref{fig:b}).  The factor $ p_{n}^8$ comes from additional redundancy if all  8 pieces of the graph are rigid (but not necessarily redundant). 
Eq.~(\ref{eq:d}) together with (\ref{eq:aaa}) provides an iterative equation for the free energy $F_{n}(p)$ resulting from (\ref{eq:aa})

\begin{equation}
F_{n}(p) = 2  N_n^{s}(p) -  N_n^{b} (p)  + R_{n} (p) .
\label{eq:e}
\end{equation}
From the eigenvalue  $\lambda_F (=8)$ for the linearized scaling of $F_{n}(p)$ at $p_c$ and large $n$, we find that the exponent $\alpha$ is  negative  (signifying a cusp) and also establish that the hyperscaling relationship to $d\nu$ is satisfied: $2-\alpha = d\nu = \ln {\lambda_F}/\ln {\lambda_1} = 3.533$.

It 
is convenient to define the number of floppy modes per degree of freedom as $f_{n}(p)= F_{n}(p)/[ 2 N_n^{s}(p)]$ so that 
$0 < f < 1$. Here $f(p)$ is the thermodynamic limit as $n \rightarrow 1$ of $f_{n}(p)$. In  Fig.~\ref{fig:f}, we show both $f$ and its second derivative with respect to $p$. 
Solving  Eqs.~(\ref{eq:e}) at the critical point gives 
$f(p_c) = 1 - 7{p_c}/6 +  {p_c^{8}}/6 = 0.00361$ 
and at small $p$, we have $f(p) = 1- 7p/6+7{p^8}/48 +...$ where the term in $p^8$ is the leading correction due to redundancy or the onset of dependent constraints.

In the 
above 
treatment  
of the rigidity percolation problem it has only been necessary to consider averages, of such things as numbers of stress-carrying bonds, redundant ones, floppy modes, etc., governed by additive composition rules.  
Such additivity is absent for processes such as percolation conductivity~\cite{RBS} (or elasticity), where probability distributions have to be rescaled.

For the additive variables of rigidity percolation, probability distributions could have been found simply (from 
algebraic recurrence relations for their Laplace transforms). These can provide further useful information e.g. for distinguishing the situations with/without central limit simplicity away from/near the transition.

To summarize,
we have shown for the first time how renormalization group procedures can be used to describe second order phase transitions involving rigidity percolation when rigidity percolates on the Berker lattice considered here. 

Outstanding questions include more rigorous approaches to establish that the number of floppy modes $F$ is the appropriate free energy for this problem. 
In addition much insight would be gained by widening the scope of the lattices covered. 

In that connection it should be mentioned that the 
Berker lattice discussed here is 
the simplest 
member of 
several families  
for which analytic results have been derived (all showing continuous transitions) which 
space precludes presenting here.
Work continues towards finding such lattices with a first order rigidity transition, and possibly a parameter to tune the transition  through a tricritical point.

We should like to thank Nihat Berker, Roger Elliott and  Sergio de Queiroz 
for useful discussions. MFT would like 
to thank Theoretical Physics at the University of Oxford for continuing summer hospitality and the National Science Foundation for support under grant  DMR 07-03973.

\end{document}